\documentclass[twocolumn, showpacs,preprintnumbers,amsmath,amssymb]{revtex4}
\usepackage{amssymb}
\usepackage{xcolor}
\usepackage{graphicx}% Include figure files
\usepackage{dcolumn}% Align table columns on decimal point
\usepackage{bm}% bold math
\usepackage{multirow}
\usepackage{booktabs}
\usepackage{natbib}
\usepackage{amssymb}
\usepackage{amsmath}	% Advanced maths commands
\usepackage{color}
\usepackage{cases} 

\newcommand{\numat}{\mbox{\boldmath $\nu$}}

\newcommand{\Smat}{\mbox{\boldmath $\cal S$}}

\newcommand{\Xmat}{\mbox{\boldmath $X$}}

\begin{document}
\preprint{APS/123-QED}

%\title{ Optical shielding of ultracold K-Cs binary collision}
\title{Electron driven reactive processes involving H$_2^+$ and HD$^+$ molecular cations in the Early Universe}

\author{E. Djuissi$^{1}$}
\author{R. Bogdan$^{2}$}
\author{A. Abdoulanziz$^{1}$}
\author{N. Pop$^{3}$}
\author{F. Iacob$^{4}$}
\author{C. Cl\'ement$^{1}$}
\author{M. D. Ep\'ee Ep\'ee$^{5}$}
\author{O. Motapon$^{5,6}$}
\author{V. Laporta$^{7}$}
\author{J. Zs Mezei$^{1,8}$}
\author{I. F. Schneider$^{1,9}$}
\affiliation{$^{1}$LOMC CNRS-UMR6294, Universit\'e le Havre Normandie, F-76058 Le Havre, France}
\affiliation{$^{2}$Dept. of Comput. \& Informat. Technol., Politehnica University Timi\c{s}oara, 300223 Timi\c{s}oara, Romania}
\affiliation{$^{3}$Department of Physical Foundation of Engineering, University Politechnica of Timisoara, 300223, Timisoara, Romania}%
\affiliation{$^{4}$Physics Faculty, West University of Timisoara, 300223,Timisoara, Romania}%
\affiliation{$^{5}$Department of Physics, Faculty of Sciences, University of Douala, P. O. Box 24157, Douala, Cameroon}%
\affiliation{$^{6}$Faculty of Science, University of Maroua, P. O. Box 814, Maroua, Cameroon}%
\affiliation{$^{7}$Istituto per la Scienza e Tecnologia dei Plasmi, CNR, 70126 Bari, Italy}
\affiliation{$^{8}$Institute for Nuclear Research (ATOMKI), H-4001 Debrecen, Hungary}%
\affiliation{$^{9}$LAC CNRS-FRE2038, Universit\'e Paris-Saclay, F-91405 Orsay, France}%
\date{\today}

\begin{abstract}
We describe the major low-energy electron-impact processes involving  $ \mathrm H ^ + _ 2 $ and $ \mathrm{HD}^+$, relevant for the astrochemistry of the early Universe: 
Dissociative recombination, elastic, inelastic and superelastic  scattering. We report cross sections and Maxwellian rate coefficients of both rotational and vibrational transitions, and outline several important features, like  isotopic, rotational and resonant effects. 
\end{abstract}

%\pacs{33.80. -b, 42.50. Hz}% PACS, the Physics and Astronomy
                             % Classification Scheme.
%\keywords{Astrochemistry, Dissociative recombination, Vibrational excitation, Early Universe}%Use showkeys class option if keyword

                              %display desired
\maketitle

\section{Introduction}

The models of the early Universe \citep{Lepp2002,Coppola2016} state that atomic hydrogen, helium and lithium - and their cations - 
have been the very first species produced by the nucleosynthetic activity which followed the Big Bang.
Later on, atoms reacted to form simple molecules, like HeH$^+$, H$_2$, HD, LiH and their cations. These latter ones face
Dissociative Recombination (DR):
\begin{equation}\label{dreq}
\mathrm{AB}^+(N_i^+,v_i^+)  + e^- \to \mathrm{A} + \mathrm{B},
\end{equation}
Ro-Vibrational Transitions (RVT):
\begin{equation}\label{ceq}
\mathrm{AB}^+(N_i^+,v_i^+) + e^-(\varepsilon) \to \mathrm{AB}^+(N_f^+,v_f^+)+ e^- (\varepsilon'), 
\end{equation}
and 
Dissociative Excitation (DE):
\begin{equation}\label{deeq}
\mathrm{AB}^+(N_i^+,v_i^+)  + e^- \to A + B^+ +e^- ,
\end{equation}
where $N_i^+/N_f^+$ and $v_i^+/v_f^+$ are the initial/final rotational and vibrational quantum numbers of the target ion, and $\varepsilon/\varepsilon'$ the  energy of the incident/scattered electron.

The dissociative recombination was first bravely proposed as elementary process in the Earth's ionosphere, as a competitor to the photoionisation providing free electrons \citep{Bates1947}. Currently, it is considered a corner-stone reaction in the synthesis of interstellar molecules and plays an important role in the ionized layers of other planets, exoplanets and their satellites. In the modeling of the kinetics of cold dilute gases, the ro-vibrational distribution of molecular species is governed by competition between formation and destruction processes, absorption, fluorescence, radiative cascades, and low-energy collisions involving neutral and ionized atomic and molecular species as well as electrons. Rate coefficients for such elementary reactions are badly needed, in particular for the chemical models of the early Universe, interstellar medium, and planetary atmospheres. 

Within a semiclassical scenario, a two-step process characterizes the  DR. First, the electron is captured by the molecular cation while exciting an electron, similarly to the dielectronic recombination of atomic cations. A neutral molecule is formed, for many species in a doubly-excited, repulsive electronic state, located above the lowest ionization potential. Second, the molecule dissociates rapidly along the potential energy curve of this dissociative state. The (re-)ejection of an electron may occur - autoionization - with low probability due to rapid dissociation which lowers the electronic energy below the lowest ionization limit. The molecule stabilizes then the electron capture by dissociating. 

The spectroscopic information, given by the output of RVT reactions (\ref{ceq}), is sensitive to the quantum numbers of the target ion (initial: \{$N_i^+$, $v_i^+$\} and final: \{$N_f^+$, $v_f^+$\}) and consequently it provides the structure of the ionized media.
 As for the RVT, they are called Elastic Collisions (EC), Inelastic Collisions and Super-Elastic Collisions (SEC) when the final energy of the electron is equal, smaller or larger respectively than the initial one.

This paper aims to illustrate our theoretical approach of the reactive collisions of electrons with  $ \mathrm H ^ + _ 2 $ and $ \mathrm{HD} ^ + $ cations at low - below 1 eV - energy, from basic ideas to  computation of  cross sections and rate coefficients, {\it via} details of the methods we use. The results we show are a part of a huge series of data we are about to produce, relevant for the kinetic modelling of the early Universe. This data generation was initiated by previous publications of our group \citep{Motapon2014, Epee2016}. In the present study, the accent is put on  $ \mathrm{HD} ^ + $ rather than on $ \mathrm H ^ + _ 2 $, since most of the latest experiments - performed in storage rings - focused on this isotopologue, is subject of quick vibrational relaxation and, consequently, more easy to vibrationally resolve than the homonuclear species. We have to notice that, whereas for this latter isotopologue, the rotational transitions involve rotational quantum numbers of strictly the same parity  (even or odd), this rule is not valid for the deuterated variant. However, we have assumed it for $ \mathrm{HD} ^ + $  too, since no data are so far available for the {\it gerade/ungerade} mixing, and since the transitions between rotational quantum numbers of different parity are much less intense than the others \citep{shafir2009}.

Our paper is organized as follows: The introduction is followed - Section \ref{ta} - by a brief description of the employed theoretical approach and of the major computational details.
In section \ref{sec:res} we present and discuss our calculated cross sections and rate coefficients. The paper ends with conclusions - Section \ref{sec:con}.

\section{Theoretical approach}\label{ta}

We presently use a stepwise version of the Multichannel Quantum Defect Theory (MQDT) to study the electron-induced reactions described in
 Eqs.~(\ref{dreq}--\ref{deeq}). In the last decade, we made evolve this approach and  applied it  successfully for computing the dissociative recombination, ro-vibrational and dissociative excitation cross sections of  H$_2^+$ and its isotopoloques~\citep{Waffeu2011,Chakrabarti2013,Motapon2014,Epee2016}, CH$^+$~\citep{Mezei2019}, SH$^+$~\citep{Kashinski2017}, BeH$^+$ and its isotopologues~\citep{Niyonzima2017,Pop2017,Niyonzima2018}, etc.

The reactive collision between an electron and a 
diatomic cation target can follow two pathways, a {\it direct} one when the electron is captured into a (most often doubly-excited) dissociation state of the neural and an {\it indirect} one where the electron is captured in a bound mono-excited Rydberg state which in turn is predissociated by the dissociative one. Both pathways involve ionization and dissociation channels, {\it open} if the total energy of the molecular system is higher than the energy of its fragmentation threshold, and 
{\it closed} in the opposite case. The open channels are responsible for the direct mechanism and for the autoionization/predissociation, while the closed ionization channels imply the electron capture into series of Rydberg states~\citep{Giusti1980, Schneider1994}. The quantum interference between the indirect and the direct mechanisms results in the {\it total} processes. 

A detailed description of method 
  has been given in previous articles~\citep{Motapon2014,Mezei2019}, and here, the main ideas and steps will be recalled.
\begin{enumerate}
\item {\it Building the interaction matrix}: Within a quasidiabatic representation, for a given set of conserved quantum numbers of the neutral system, $\Lambda$ (projection of the electronic angular momentum on the internuclear axis) and $N$ (total rotational quantum number), the interaction matrix is based on the couplings between ionization channels - associated with the ro-vibrational levels $N^+,v^+$ of the cation and with the orbital quantum number $l$ of the incident/Rydberg electron - and dissociation channels.
\item {\it Computation of the reaction matrix}: We adopt the second-order perturbative `exact' solution~\citep{Ngassam2003} of the Lippmann-Schwinger integral equation~\citep{Florescu2003,Motapon2006}.
\item {\it Diagonalization of the reaction matrix}: We end up in building the {\it eigenchannel} short-range representation.
\item {\it Frame transformation} We switch from the {\it Born-Oppenheimer} (short-range) representation, characterized by $N$, $v$, and $\Lambda$ to the close-coupling (long-range) representation, characterized by $N^+$, $v^+$, $\Lambda^+$ for the ion, and $l$ (orbital quantum number) for the incident/Rydberg electron.
\begin{figure}[t]
\begin{center}
\includegraphics[width=0.5\textwidth]{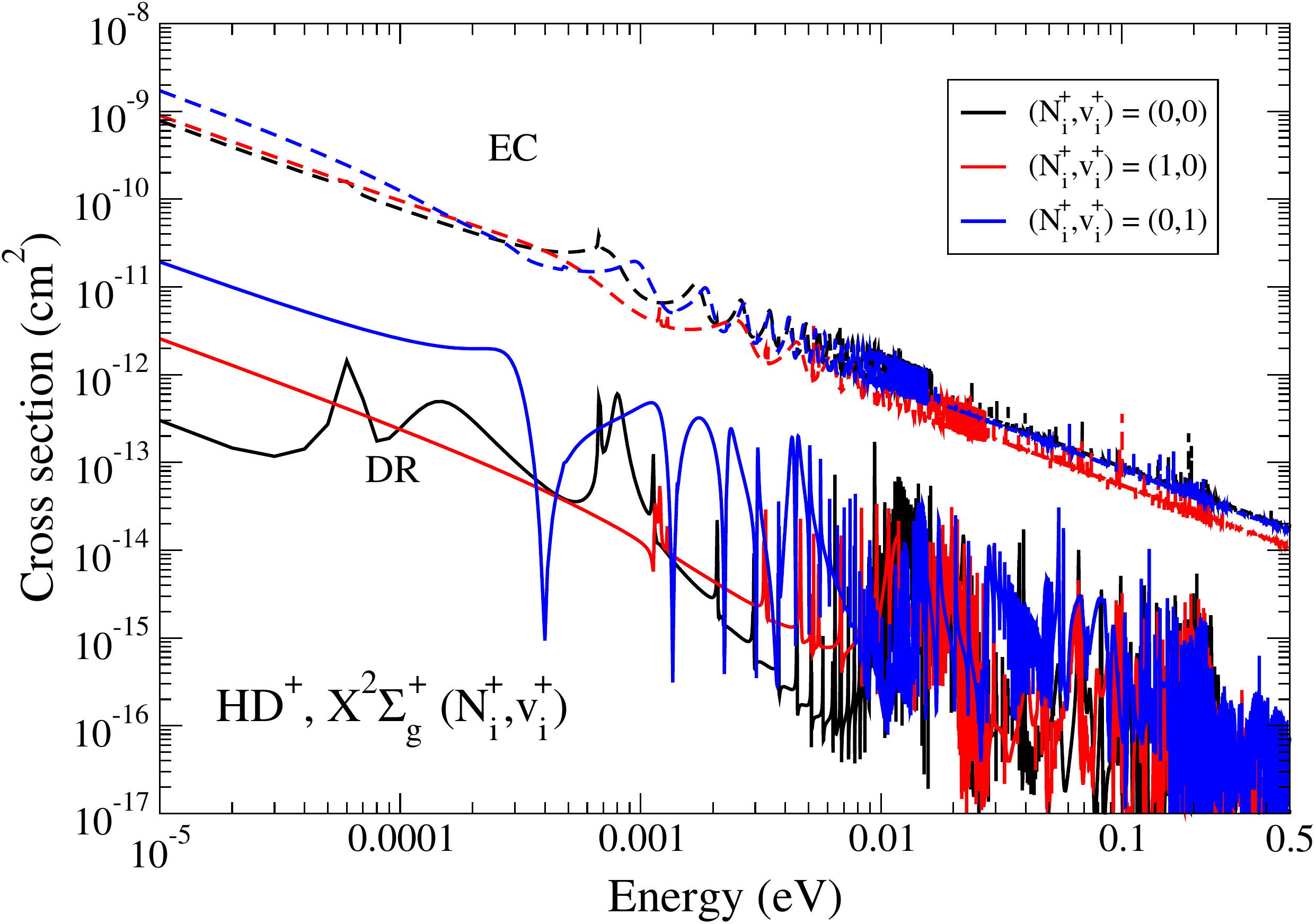}
\caption{
Dissociative Recombination (DR) and Elastic Collisions (EC)
of HD$^{+}($X$^{2}\Sigma_{g}^{+})$, 
effect of the excitation of the target.
Black: target in its ground state ($N_{i}^{+}=0,  v_{i}^{+} = 0 $).
Red: target rotationally excited ($N_{i}^{+}=1,  v_{i}^{+} = 0 $).
Blue: target vibrationally excited ($N_{i}^{+}=0,  v_{i}^{+} = 1 $).
}
 \label{fig:1}
\end{center}
\end{figure}
\item {\it Building of the generalized scattering matrix}: Based on the frame-transformation coefficients and on the Cayley transform, we obtain the generalized scattering matrix, organized in blocks associated with energetically open ($o$) and/or closed ($c$) channels:
\begin{equation}
\Xmat=
\left(\begin{array}{cc} \Xmat_{oo} & \Xmat_{oc}\\
                   \Xmat_{co} & \Xmat_{cc} \end{array} \right).
\end{equation}
\item {\it  Building of the physical scattering matrix}: Applying the method of~"elimination of the closed channels"~\citep{Seaton1983} we get the scattering matrix:
\begin{equation}\label{eq:solve3}
\Smat=\Xmat_{oo}-\Xmat_{oc}\frac{1}{\Xmat_{cc}-\exp(-i2\pi \numat)}\Xmat_{co}.
\end{equation}
The diagonal matrix {\numat} in the denominator above
contains 
the effective quantum numbers corresponding to the vibrational thresholds of the 
closed ionisation channels  at the current total energy of the system. 
\begin{figure*}[t]
\includegraphics[width=\textwidth]{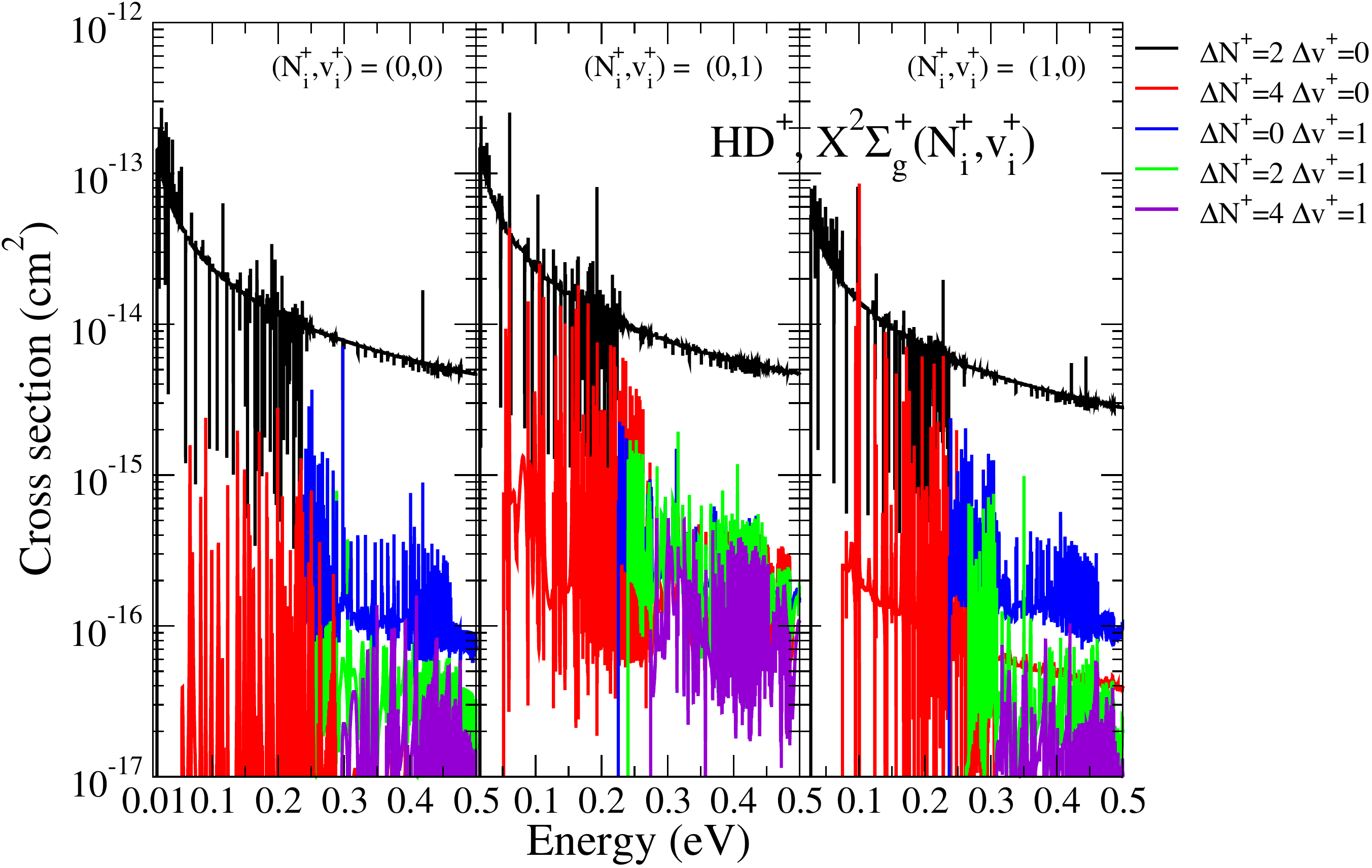}
\caption{Ro-Vibrational Excitation (RVE)
of HD$^{+}($X$^{2}\Sigma_{g}^{+})$,
effect of the excitation of the target.
Left: target in its ground state ($N_{i}^{+}=0,  v_{i}^{+} = 0 $).
Middle: target vibrationally excited ($N_{i}^{+}=0,  v_{i}^{+} = 1 $).
Right: target rotationally excited ($N_{i}^{+}=1,  v_{i}^{+} = 0 $).}
\label{fig:2}
\end{figure*}

\item{\it Computation of the cross-sections}: For the target initially in a state characterized  by the quantum numbers $N_i^+,v_i^+,\Lambda_i^+$, and for the energy of the incident electron $\varepsilon$, the dissociative recombination and the ro-vibrational transitions 
global cross-sections read respectively 
\begin{eqnarray}\label{eqDR1}
\sigma _{diss \leftarrow N_{i}^{+}v_{i}^{+}} &=&
		\sum_{\Lambda,sym} \sigma _{diss \leftarrow N_{i}^{+}v_{i}^{+}}^{(sym,\Lambda)},\\
\sigma _{diss \leftarrow N_{i}^{+}v_{i}^{+}}^{(sym,\Lambda)} &=& 
%		\frac{\pi}{4\varepsilon} \rho^{(sym,\Lambda)} \sum_{N} \frac{2N+1}{2N_{i}^{+}+1} \sum_{l,j}\mid S^{(sym,\Lambda,N)}_{d_{j},N_{i}^{+}v_{i}^{+}l}\mid^2,\\\label{eqDR2}
		\frac{\pi}{4\varepsilon} \rho^{(sym,\Lambda)} \sum_{N} \frac{2N+1}{2N_{i}^{+}+1} \times \\\nonumber
		&\times&\sum_{l,j}\mid S^{(sym,\Lambda,N)}_{d_{j},N_{i}^{+}v_{i}^{+}l}\mid^2,\\\label{eqDR2}
\sigma _{N_{f}^{+}v_{f}^{+} \leftarrow N_{i}^{+}v_{i}^{+}} &=&
		\sum_{\Lambda,sym} \sigma _{N_{f}^{+}v_{f}^{+}\leftarrow N_{i}^{+}v_{i}^{+}}^{(sym,\Lambda)},\\\label{eqVE_VdE1}
\sigma _{N_{f}^{+}v_{f}^{+}\leftarrow N_{i}^{+}v_{i}^{+}}^{(sym,\Lambda)} &=&
		\frac{\pi}{4\varepsilon} \rho^{(sym,\Lambda)} \sum_{N} \frac{2N+1}{2N_{i}^{+}+1} \times \\\nonumber
		&\times&\sum_{l,l'}\mid S_{N_{f}^{+} v_{f}^{+}l',N_{i}^{+} v_{i}^{+}l}^{(sym,\Lambda,N)}-\delta_{N_{i}^{+}N_{f}^{+}}\delta_{v_{i}^{+}v_{f}^{+}}\delta_{ll'}\mid^2,
\label{eqVE_VdE2}
\end{eqnarray}
\noindent where $sym$ is refering to the inversion symmetry  - gerade/ungerade - and to the spin quantum number of the neutral system,  $N$ is standing for its total rotational quantum number, and $\rho^{(sym,\Lambda)}$ is the ratio between the multiplicities of the neutral system and of the ion.
\end{enumerate}

\section{Results and discussions}\label{sec:res} 

%\begin{sidewaysfigure}[t]
%\end{sidewaysfigure}

The results presented in this work are the first ones going beyond our previous studies performed on HD$^{+}~$~\citep{Waffeu2011,Motapon2014} and H$^+_ 2$~\citep{Epee2016} for low collision energies relevant for astrophysical applications. Indeed, not only we extended the range of the incident energy of the electron  but, furthermore, we considered for the first time simultaneous rotational and vibrational transitions (excitations and/or de-excitations).  

The calculations were performed using the step-wise MQDT method including rotation, briefly outlined in the previous section. We have used the same molecular structure data sets as those from our previous studies~\citep{Waffeu2011,Motapon2014,Epee2016}. The cross sections have been calculated with the inclusion of both {\it direct} and {\it indirect} mechanisms for the $\Sigma^+$, $\Pi$  and $\Delta$ - singlet and triplet, gerade and ungerade - symmetries at the highest (second) order of perturbation theory. The energy range considered here was $10^{-5} - 1.7$ eV, while the energy step was taken as $0.01$ meV. 

\begin{figure}[t]
\begin{center}
\includegraphics[width=0.5\textwidth]{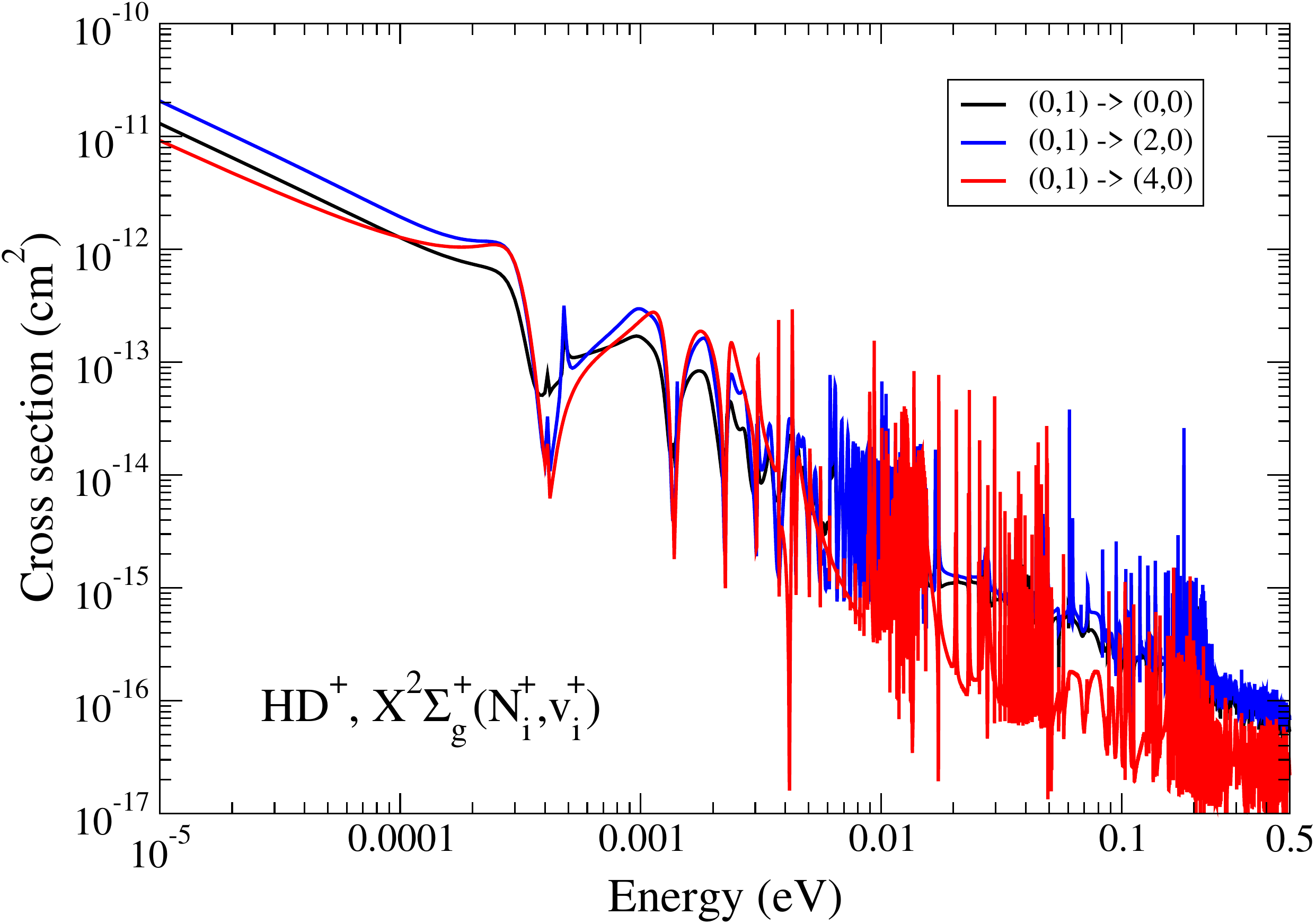}
\caption{De-excitation
 of the lowest vibrationally-excited level ($N_{i}^{+}=0,  v_{i}^{+} = 1 $) 
 of HD$^{+}($X$^{2}\Sigma_{g}^{+})$, 
 dependence on the {\it final} ro-vibrational state:  
 ($N_{i}^{+}=0,  v_{i}^{+} = 0 $) - vibrational de-excitation, black,  
 ($N_{i}^{+}=2,  v_{i}^{+} = 0$) and ($N_{i}^{+}=4,  v_{i}^{+} = 0$) -   
rotational excitation and vibrational de-excitation, blue and red respectively.}
\label{fig:3}
\end{center}
\end{figure}

\begin{figure}[t]
\begin{center}
\includegraphics[width=0.5\textwidth]{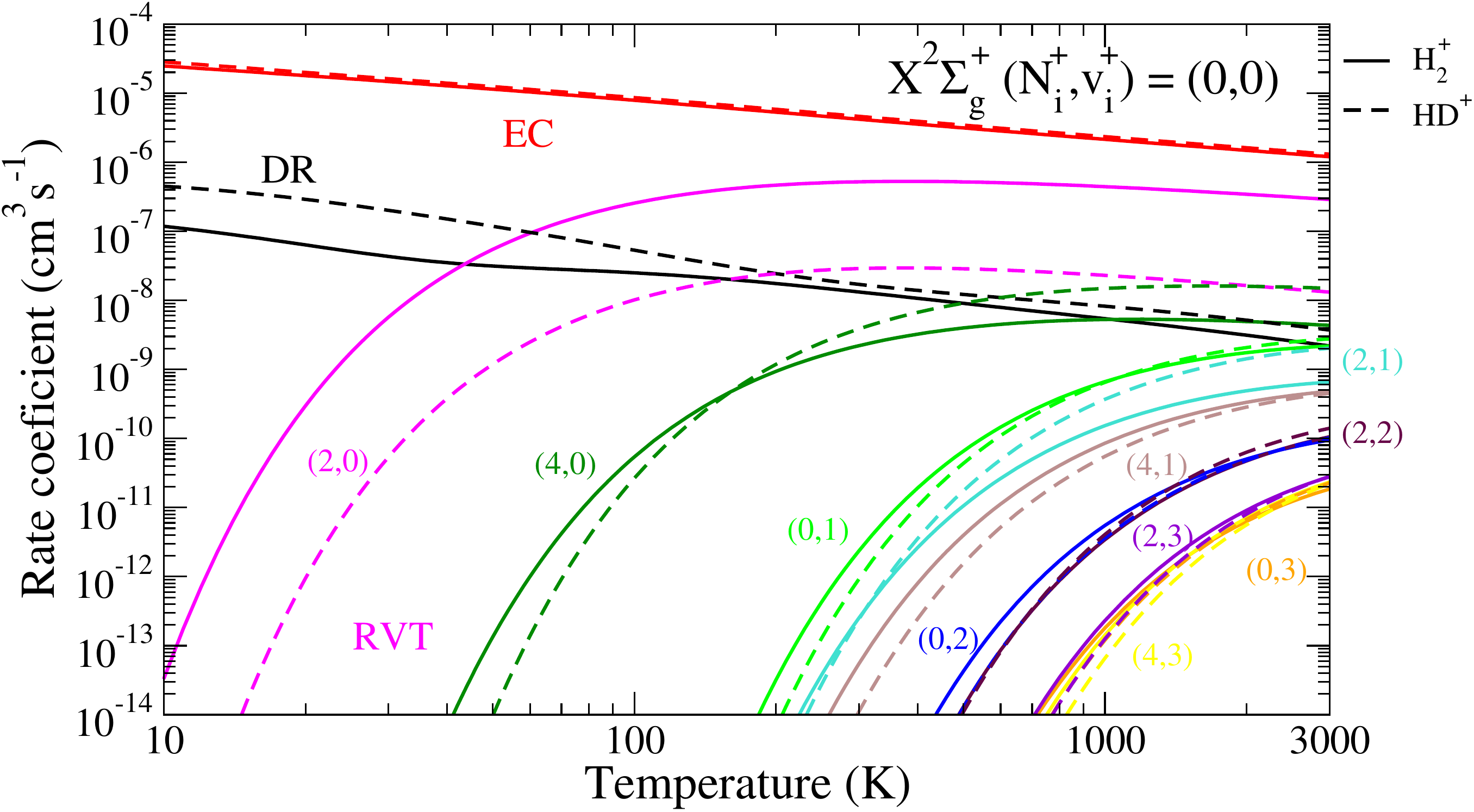}
\caption{
Electron-impact dissociative recombination and ro-vibrational transitions of  
$ \mathrm{HD} ^ + $ and $ \mathrm H ^ + _ 2 $
in their ground state: Maxwell rate coefficients.} 
\label{fig:4}
\end{center}
\end{figure}

Our results are presented in figures~\ref{fig:1}-\ref{fig:4}, where we have chosen three different initial ro-vibrational levels of the ground electronic state of both target systems, namely $(N_i^+,v_i^+)=(0,0)$, $(1,0)$ and $(0,1)$, 
corresponding to {\it ground} state, lowest rotationally {\it excited} state and lowest {\it vibrationally} excited state respectively.

The calculated cross sections (Figs.~\ref{fig:1}-\ref{fig:3}) are restricted to low collision energies - $\varepsilon\leq0.5$ eV - where the rotational effects are the most relevant, while for calculating the rate coefficients (Fig.~\ref{fig:4}) we have used the cross sections on the whole energy range.

Figure~\ref{fig:1} shows the dissociative recombination (solid lines) and resonant elastic scattering (dashed lines) cross sections of HD$^+$ for the previously defined three initial target states. 
The background $\frac{1}{\varepsilon}$ trend is due to the direct mechanism, while the resonant structures correspond to the temporary captures of the incident electron into ro-vibrational levels of  
Rydberg states - indirect mechanism. 

The cross section of EC exceeds the DR by at least two orders of magnitudes, and we found that the importance of resonances and of the target-excitation effects are much less pronounced than in the case of the DR, both on order of magnitude and on position and number density of resonances. 

Figure~\ref{fig:2} illustrates the dependence of the ro-vibrational excitation - $\Delta N^+=0,2,4$ and $\Delta v^+=0,1$ - cross section on the excitation of the  target. 
 While the most prominent $\Delta N^+=2$ $\Delta v^+=0$ transition shows very little dependence, those involving more change in the ro-vibrational state are more sensitive to the target state.
 The largest sensitivity corresponds to the $\Delta N^+=4$ transition, where a unity change in either and/or both rotational and vibrational quanta increases the cross section with more than one order of magnitude. 
The cross sections display
threshold effects and, 
imilarly to DR, prominent resonances. 
On the other hand, concerning the intensity of the transitions, the  $\Delta N^+=2$  $\Delta v^+=0$ one
is followed by the $\Delta v^+=1$ purely vibrational (blue curves) and by either the $\Delta N^+=4$ purely rotational (red curves) or $\Delta N^+=2,\,\,\Delta v^+=1$ (green curves) 
 'mixed' ro-vibrational excitations. The smallest cross sections caracterize
 the $\Delta N^+=4,\,\,\Delta v^+=1$ excitations. 

Figure \ref{fig:3} is an illustration of the dependence of the de-excitation cross section on
the {\it final} ro-vibrational state of the target ion. We have chosen as example the case of 
HD$^{+}($X$^{2}\Sigma_{g}^{+})$  initially on its ($N_{i}^{+}=0,  v_{i}^{+} = 1 $) level.
The black, blue and red curves correspond to the 
($\Delta v^{+}=-1$,  $\Delta N^{+}=0)$, 
($\Delta v^{+}=-1$,  $\Delta N^{+}=2)$, 
and 
($\Delta v^{+}=-1$,  $\Delta N^{+}=4)$
transitions respectively. 
While bellow $10$ meV of collision energy the three cross sections have roughly the same shape and magnitude, above $10$ meV the $\Delta N^{+}=4$ transition shows quite different resonance patterns and becomes almost one order of magnitude smaller than the other two transitions.

And finally, in order to obtain the thermal rate coefficients, we have convoluted our cross sections with the isotropic Maxwell distribution function for the kinetic energy of the incident electrons:

\begin{equation}\label{rate}
\alpha(T)=\frac{8\pi}{{(2\pi kT)}^{3/2}}\int_{0}^{+\infty}\sigma(\varepsilon)\varepsilon\exp(-\varepsilon/kT)d\varepsilon,
\end{equation}

\noindent where $\sigma$ is one of the cross sections calculated according the eqs. (\ref{eqDR1}) and (\ref{eqVE_VdE1}) and $k$ is the Boltzmann constant.

The 
DR and RVT Maxwell rate coefficients 
for electron temperatures between 10 and 3000 K
are given in Fig.~\ref{fig:4} for HD$^+$ and H$_2^+$ targets initially in their ground ro-vibrational state 
$(N_i^+,v_i^+)=(0,0)$.
The resonant EC rates are overall constantly the highest. Among the othess proceses the DR ones predominates at very low temperatures, and is surpassed by the lowest rotational excitation above 50 K. The higher excitations become notable above 2000 K only. 
Figure~\ref{fig:4} also illustrates the isotopic effects: it is important for $\Delta N^{+}=2$ and $4$ rotational excitations and for DR below $100$ K electron temperature. 

\section{Conclusions}\label{sec:con}
The multichannel quantum defect theory results in the detailed modeling of the reactive collisions of electrons with H$_2^+$ and HD$^+$. The account of the interfering mechanisms - direct and indirect - as well as of the major interactions - Rydberg/valence, ro-vibronic and rotational - result in accurate state-to-state theoretical cross sections and rate coefficients. 

% and molecular cations.% provided that the molecular structure of the target and of the neutral complex has been explored and quantitatively characterized by quantum chemistry and R- matrix methods.
 %High accuracy has been achieved for the H$_2^+$/HD$^+$ benchmark systems, and the %progressive account of the numerous mechanisms, and interactions like Rydberg-valence %couplings and rotational effects result in theoretical cross sections and rate coefficients in %increasing agreement with the measurements in storage rings.

The results of the present paper are a first step in the extension of our previous studies on reactive collision 
of HD$^{+}$ and H$^+_ 2$ with electrons to a wider range of incident collision energy and to mixed - i.e. simultaneous rotational and vibrational - transitions. 
The results concerning higher ro-vibrational levels are the subject of an ongoing work. We do expect a strong dependence of the cross sections and Maxwell rate coefficients on the target state. The provided collisional data are available on demand to be used in the kinetics modeling in astrochemistry - early Universe, interstellar molecular space - and cold plasma physics.

%completes a series of studies performed on HD$^{+}$ and H$^+_ 2$ targets by extending the %domain of the incident collision energy range of the electron and by considering for the first time %rotational and vibrational transitions (excitations and/or de-excitations) simultaneously. 

%{\color{magenta}hereIstopped}

%{\color{green} 2do:

%2) mentioning that HD is treated as H2: gerade/ungerade, Delta N+=1,3,  are assumed to be negligible.

%}

\begin{acknowledgements}
		The authors acknowledge support from
	Agence Nationale de la Recherche {\it via} the 
project MONA,
	Centre National de la Recherche Scientifique {\it via} the 
GdR TheMS,
PCMI program of INSU (ColEM project, co-funded by CEA and CNES),
PHC program Galil\'ee between France and Italy, and
DYMCOM project,
	F\'ed\'eration de Recherche Fusion par Confinement Magn\'etique (CNRS, CEA and Eurofusion),
	La R\'egion Normandie, FEDER and LabEx EMC$^3$ {\it via} the projects 
Bioengine, 
EMoPlaF, 
CO$_2$-VIRIDIS and 
PTOLEMEE,
	COMUE Normandie Universit\'e, 
	the Institute for Energy, Propulsion and Environment (FR-IEPE), 
	the European Union {\it via}
COST (European Cooperation in Science and Technology) actions 
	TUMIEE (CA17126), 
 	MW-GAIA (CA18104)
and 
	MD-GAS (CA18212),
and 
	ERASMUS-plus conventions between Universit\'e Le Havre Normandie and Politehnica University Timisoara, West University Timisoara and University College London.
		NP is grateful for the support of the Romanian Ministry of Research and Innovation, project no. 10PFE/16.10.2018, PERFORM-TECH-UPT. 
		JZsM thanks the financial support of the National Research, Development and Innovation Fund of Hungary, under the FK 19 funding scheme with project no. FK 132989. 
		IFS and VL thank Carla Coppola and Daniele Galli for having suggested the present work, as well as for their constant interest and encouragement.

\end{acknowledgements}
%%%%%%%%%%%%%%%%%%%% REFERENCES %%%%%%%%%%%%%%%%%%

%\section*{Data availability}
%The data underlying this article will be shared on reasonable request to the corresponding author.

\end{document}